# Machine Learning Pipeline for Denoising Low Signal-To-Noise Ratio and Out-of-Distribution Transmission Electron Microscopy Datasets

Brian Lee[1], Meng Li[1], Judith C. Yang,[1,2,*] Dmitri N. Zakharov,[1,*] Xiaohui Qu[1,*]

[1]Center for Functional Nanomaterials, Brookhaven National Laboratory, Upton, NY, United States
[2]Department of Chemical and Petroleum Engineering, University of Pittsburgh, Pittsburgh, PA, United States
*Corresponding author email: jyang1@bnl.gov, dzakharov@bnl.gov, xiaqu@bnl.gov

## Abstract

High-resolution transmission electron microscopy (HRTEM) is crucial for observing material's structural and morphological evolution at Angstrom scales, but the electron beam can alter these processes. Devices such as CMOS-based direct-electron detectors operating in electron-counting mode can be utilized to substantially reduce the electron dosage. However, the resulting images often lead to low signal-to-noise ratio, which requires frame integration that sacrifices temporal resolution. Several machine learning (ML) models have been recently developed to successfully denoise HRTEM images. Yet, these models are often computationally expensive and their inference speeds on GPUs are outpaced by the imaging speed of advanced detectors, precluding in situ analysis. Furthermore, the performance of these denoising models on datasets with imaging conditions that deviate from the training datasets have not been evaluated. To mitigate these gaps, we propose a new self-supervised ML denoising pipeline specifically designed for time-series HRTEM images. This pipeline integrates a blind-spot convolution neural network with pre-processing and post-processing steps including drift correction and low-pass filtering. Results demonstrate that our model outperforms various other ML and non-ML denoising methods in noise reduction and contrast enhancement, leading to improved visual clarity of atomic features. Additionally, the model is drastically faster than U-Net-based ML models and demonstrates excellent out-of-distribution generalization. The model's computational inference speed is in the order of milliseconds per image, rendering it suitable for application in in-situ HRTEM experiments.

# 1. Introduction

Advancements in high-resolution transmission electron microscopy (HRTEM) have enabled observations of Angstrom length and millisecond time scale structural and morphological evolution of materials governing phenomena such as oxidation-reduction reactions,[1-7] catalytic reactions,[8-11] nanoparticle fluxionality,[12-14] and biomolecular processes.[15] For example, recent studies have demonstrated that surfaces of platinum nanoparticles transition from stable to metastable structures at millisecond scales[12] and this fluxionality showed correlation with CO oxidation activity in Pt catalysts[13]. In situ HRTEM studies on copper oxidation have demonstrated an intermittent, stop-and-go growth mechanism that involves reorganization of the surface structure of copper,[16] contrasting traditional theory that assumes homogeneous oxide growth.[17] Such findings have led to improved theoretical and practical understanding, as well as optimization of these processes.

Despite these successes, utilizing HRTEM involves a fundamental dilemma, where the electron beam used to detect the atomic-scale events can alter, accelerate, or suppress the observed nanoprocesses.[18, 19] CMOS-based direct-electron detectors[20] (DED) can mitigate this problem by operating in electron-counting mode, which preserves high detective quantum efficiency even when the electron dose is significantly reduced.[21] However, such low doses produce severely noisy images whose noise characteristics can vary significantly depending on the detector type and acquisition mode. A common workaround is to integrate frames over time to increase the signal-to-noise ratio, but this approach results in loss of temporal resolution. Reducing the noise without sacrificing the temporal or spatial resolutions is therefore essential for exploiting the full potential of in situ HRTEM experiments.

To mitigate the noise in HRTEM data, many machine learning (ML) models have been developed. Several studies[22-24] developed supervised ML models that are trained to convert noisy input images into clean output images. For example, Ihara et al.[22] collected slow-scan, high-quality reference images alongside fast-scan, noisy frames of the same regions and trained a U-Net[25] to map the latter onto the former. Lobato et al.[23] utilized a generative adversarial network that cleans synthetically corrupted images. However, such ground-truth references are often unattainable for in situ HRTEM, requiring unsupervised or self-supervised models.[26-30]

For self-supervised models, blind spot convolutional neural networks (CNN) are utilized in which the denoised pixels are predicted by ML models without observing the noisy pixels at the location. These models assume that the underlying signals of pixels are correlated with their neighboring pixels while the noises are independent of their neighbors. One notable blind-spot model is the Noise2Void model[27] which predicts noise while masking a fraction of the input images. Another breakthrough model for HRTEM is unsupervised deep video denoising (UDVD) model developed by Sheth et al[12, 31], which utilizes the framework proposed by Laine et al.[32] This model creates four rotated (0°, 90°, 180°, 270°) copies of the input image and utilizes separate U-Nets that are causal in one direction by shifting the input pixels, hence not observing the central pixels. Although this strategy produces impressive restorations, the computational cost remains high because of the need of data duplication and large number of parameters associated with U-Net. In addition, its performance on datasets with different distribution of pixels compared to the training datasets are not evaluated. It is not rare that a neural network learns spurious correlations in the training set, which are unstable across environments and fails disastrously to images different from the training set. This is especially true for HRTEM as every camera runs at vastly different operating modes and conditions. Accordingly, ML models that can generalize to out-of-distribution (OOD)[33, 34] datasets are highly demanded for applications such as analysis of in situ HRTEM experiments.

In this paper, we develop a comprehensive ML pipeline to address the challenges of denoising time-series HRTEM, focusing on OOD dataset generalization and computational efficiency. Our approach integrates a self-supervised denoising ML model with pre-processing (drift correction) and post-processing (low-pass filtering) steps. For the denoising ML model, unlike the pixel-centric strategies of Noise2Void[27] or UDVD[31] that directly operates on the input pixel data, we adopt a weight-centric approach where the weight of the central kernel of the convolution filter is fixed to zero using a specialized backpropagation procedure. Such framework, proposed by Wu et al,[35] has shown good denoising performance on synthetic and real photography.[35, 36] We find that the our model performs significantly better compared to publicly available ML models and non-ML denoising methods in both noise reduction and contrast enhancement, improving the visual clarity of atomic features especially for OOD datasets. Computational inference speed of the model reached hundreds of frames per second for typical HRTEM images, enabling its application to in situ HRTEM experiments.

## 2. Results

**2.1 Weight Centric Denoising Model Architecture**

The ML model for this study utilizes a blind-spot kernel design to denoise HRTEM images in self-supervised manner. The model consists of three parts depicted in Fig. 1. First, we use a two-dimensional image denoising model that we denote as weight-centric image denoising model (WCID). The purpose of this model is to provide preliminary denoising so drift-correction algorithms can be applied to noisy time-series HRTEM images with significant drift. WCID may be omitted if there is no drift or if the images can be drift-corrected without the model. Second, we utilize a three-dimensional video denoising model called weight-centric video denoising model (WCVD) on the drift corrected time-series HRTEM images. Finally, a low-pass filter is applied to the output of the WCVD model to obtain the denoised video. We note that the WCID and WCVD have equivalent ML architectures except that the kernels of convolutional neural networks (CNN) are either two or three dimensional.

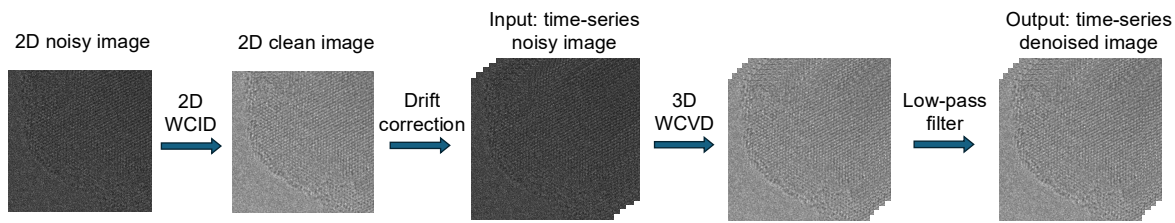

**Figure 1. Schematics of the denoising pipeline.** Denoising pipeline consists of drift correction, blind-spot denoising ML model, and low-pass filter. Drift correction is applied following WCID that provide preliminary denoising to facilitate drift correction. The WCVD and WCID share same ML architecture except for the dimensions of the CNN kernels.

WCID and WCVD are self-supervised denoising models with the architectures depicted in the Supplementary Materials (SM) Fig. S1. The HRTEM images/videos are passed through a single layer of 2D/3D blind convolution kernel (BCK, depicted in Fig. S2a), where the weight and the gradient of the central pixel is fixed to 0. BCK does not have bias. We employ BCK under the assumption that the HRTEM images are combinations of the signals from the nanomaterials, whose pixel values are spatially correlated with their neighboring pixel values, and the noise that is not spatially correlated. Therefore, BCK allows us to train kernel weights that suppress the spatially uncorrelated noise and enhance the spatially correlated signals that can be inferred from the neighboring pixels.

The outputs of BCK are then passed through $1 \times 1 \times 1$ CNN layers followed by leaky rectified linear unit (ReLU) activation function for the middle layers and ReLU activation functions for the last layer to add non-linearity to the models. For the last step of denoising, we apply a low-pass filter to the images after fast Fourier transform (FFT) operations to remove the high frequency noise. Detailed description of the low-pass filter is given in the SM section 2. The objective functions for these models are to minimize the mean squared error between the input images/videos and the model outputs. More detailed descriptions of the models and the training procedures are provided in the Method section 4.1.

**2.2 In-distribution and out-of-distribution datasets**

For training and testing the denoising models, we employ three datasets denoted as Pt, Au, and Co depicted in Fig. 2 and Fig. S5. The Pt dataset is from Crozier et al.[12, 37] and depicts platinum nanoparticle (NP) supported on $CeO_2$. The scales of the images were inferred as 0.0067 nm/pixel based on the atomic diameter of Pt (0.25 nm noted in the supplementary materials of Crozier et al.) and the average diameter of atoms in the dataset (38.7 pixels). The Au and Co datasets were HRTEM images obtained in-lab. The Au dataset consists of a single gold NP with diameter of 5 nm with scale of 0.0169 nm/pixel. The Co dataset was cobalt on $CeO_2$ with scale of 0.0209 nm/pixel. The Pt and Au datasets were observed in electron counting mode with low electron dose rates, while Co datasets were from linear integration mode. For each dataset, we use 20 time-series images. The spatial dimensions of Au and Co datasets were 512×512 pixels while the spatial dimension of Pt was 1024×1024 pixels. Detailed description of the datasets is given in the Method section 4.2.

As the datasets consisted of diverse materials, imaging conditions, and magnifications, the image qualities as well as the pixel intensities varied significantly. As depicted in the histogram of intensities in Fig. S5 of Supplementary Materials, the pixel intensities for Pt datasets varied from 0-255 as the images have been exported into TIFF files. The intensities for Au varied from 0-5, representing the raw data of the electron counting modes. Both Pt and Au dataset have quantized pixel values in the histogram due to the electron counting modes. The Co datasets displayed a distribution of intensities of 100~1000. Due to the difference in scales, the atomic column diameters were ~ 40 pixels for the Pt dataset and ~10 for Au and Co datasets.

**2.3 Denoising model performance**

The WCVD denoising model was trained only on the in-distribution (ID) Pt dataset and tested on out-of-distribution (OOD) datasets of Au and Co nanoparticle (NP) samples. To evaluate the model performance, we compared our results with those obtained from non-ML denoising techniques, such as Gaussian filtering, and ML models such as UDVD[12], autoencoder[38, 39], and Noise2Void.[27] All models were tested on drift-corrected images. For non-ML denoising, the Gaussian 2D refers to Gaussian filtering on individual slices while Gaussian 3D applies Gaussian filtering to 3D tensors of the HRTEM images including both spatial and temporal dimensions. The UDVD is a denoising ML models developed for electron microscopy and incorporate U-Net architectures. This model was not retrained and the published weights were utilized. The Noise2Void, autoencoder, and WCVD were trained on the Pt dataset. Therefore, the Au and Co datasets were OOD for all ML models while Pt was the ID dataset. More descriptions of these methods are provided in the Method section 4.3.

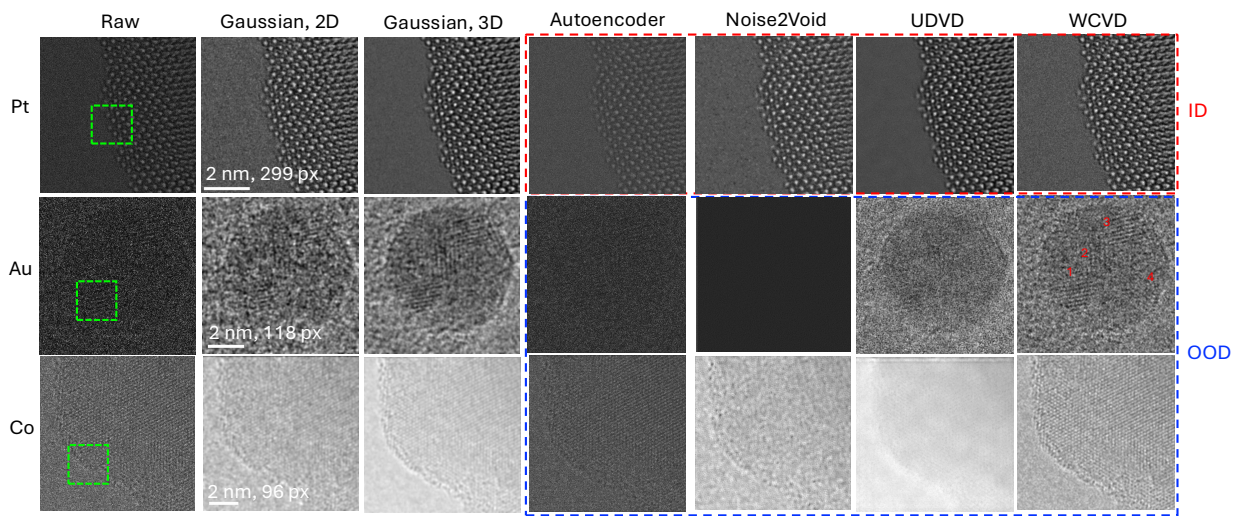

**Figure 2. Raw and denoised images.** The raw, gaussian-filtered, and ML denoised (autoencoder, Noise2Void, UDVD, WCVD) example images of Pt, Au, Co datasets are depicted. The Pt dataset is publicly available,[37] while Au and Co were taken in-lab. The red rectangles represent the dataset the ML models were trained on (in distribution, ID), while the blue rectangles represent the out-of-distribution (OOD) datasets. The green rectangles represent the magnified regions in Fig. 3.

The resulting comparisons are depicted in Fig. 2, with larger magnitude images depicted in SM Fig. S6, S7, and S8. Visual inspection of the results demonstrates that all denoising methods generally enhance the contrasts of the images around atomic columns and suppress noises that are prevalent for both the gas and NP. However, the results significantly differ between the ID (Pt) dataset and OOD (Au, Co) datasets. For Pt, all methods lead to distinct atomic columns and interfaces. On the other hand, while the Gaussian 3D and WCVD lead to discernable atomic columns, the other ML models lead to generally blurry image qualities and undiscernible features.

Several factors may contribute to such difference. First, the Pt images are taken at higher magnification with atom diameters ~ 40 pixels while the atomic diameters for Au and Co datasets were ~ 10 pixels, leading to significant domain difference beyond the pixel intensity distributions shown in Fig. S5. Second, the ML architecture of WCVD utilizes a single layer to gather information from the neighboring pixels for a particular pixel-level prediction. However, other ML models utilize deep CNN layers such as U-Nets that utilizes multiple CNN at varying levels of dimensionalities in the latent space. Such design may lead to leakage of the central pixel information and lead to overfitting. In addition, WCVD has orders of magnitude lower number of parameters, further reducing the possibility of overfitting. Such characteristics may lead the other ML models to overfit to ID datasets while the WCVD flexibly handles the OOD datasets.

In addition, the difference between Gaussian 2D and Gaussian 3D filtered results suggest that the temporal information is crucial for the highly noisy Au dataset. Specifically, the denoised Au snapshot for Gaussian 3D and WCVD depicts clearly 4 areas with resolved lattices (see Fig. 2 WCVD panel for Au), whereas the areas #1 and #4 are not distinctly resolved for the image from Gaussian 2D set. For the less noisy Pt and Co samples, the difference between Gaussian 2D and Gaussian 3D are not as distinct. While UDVD utilizes temporally adjacent frames (5 frames total) as input, the inputs for the Noise2Void and autoencoder incorporate only spatial information. Such results suggest that incorporation of the temporal dimension can be crucial for low signal data that are often obtained from low electron dose rate TEM images with high spatial and temporal resolutions.

We further evaluate the model performance by examining the magnified view of the areas marked by green rectangles in Fig. 2. Fig. 3 display magnified images and line scan over atomic columns. The results from Gaussian 3D, UDVD, and WCVD with raw data as reference are depicted here and the results for the other models are in Fig. S9 of SM. The line scans were taken along atomic columns by averaging the pixel intensities along a line with width of 25 pixels. As was observed in Fig. 2, all denoising models displayed distinct atomic lattices for Pt images. Line scans of Pt atom show that the peak and valley positions of the pixel intensities were at the same locations for Gaussian 3D, UDVD, and WCVD models, which is also apparent in the raw dataset. The contrast ratios were similar as well and the noise observed from the raw data demonstrated by the jagged spikes in the intensities were removed.

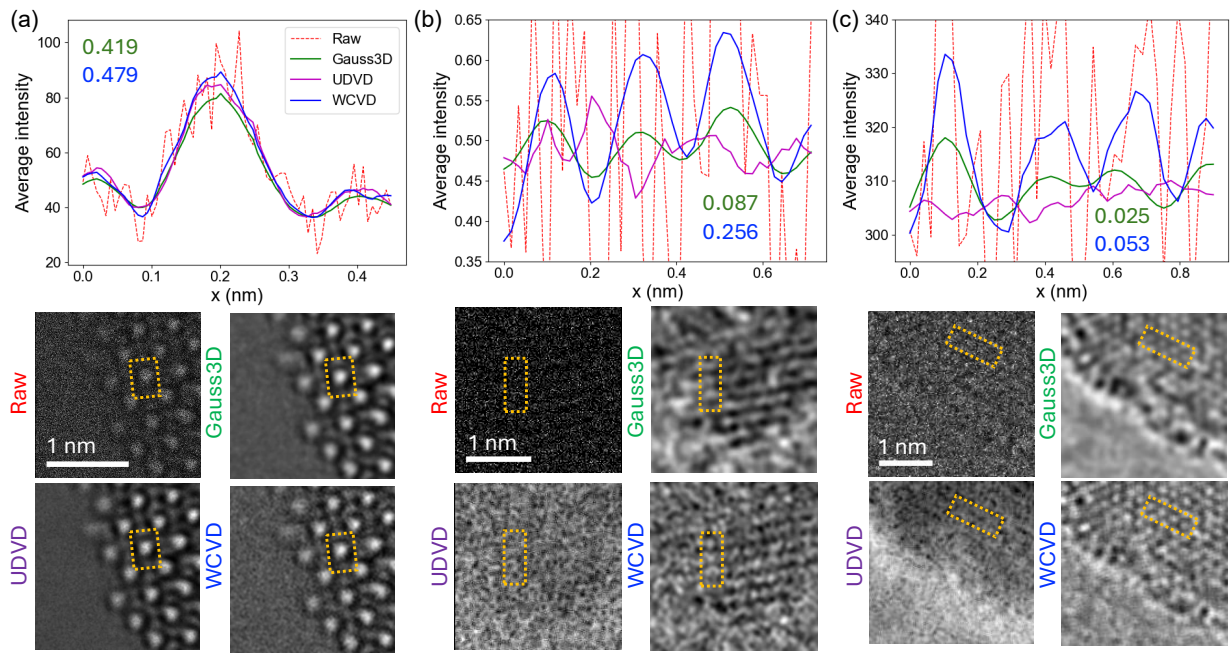

**Figure 3. Magnified regions and line scans of the atomic columns.** The datasets depicted are (a) Pt, (b) Au, and (c) Co and the denoising methods are Gaussian 3D (green), UDVD (magenta), and WCVD (blue). Raw data results are depicted in red. The line scan areas are marked by yellow rectangles in the magnified images. The numbers on the line scan plots correspond to the image contrasts for the WCVD (blue) and Gaussian 3D (green).

For the OOD datasets of Au and Co, the line scans of the raw data revealed that peaks and valleys of the intensities corresponding to the atomic lattices are not apparent without denoising, significantly differing from the Pt dataset. In addition, the peaks for OOD datasets are more proximate to each other (~10 pixels) compared to the Pt peaks (~40 pixels) because of higher magnification. Such differences in the signals of the nanomaterials of the three datasets, in addition to the differences in the noise characteristics, may contribute to the poor performance of the denoising models except WCVD and Gaussian 3D.

While all denoising models removed the spikes in intensities, all ML denoising models except WCVD did not clarify the atomic columns with clear peaks and valleys. For the Gaussian 3D and WCVD, the locations of the peaks and valleys were similar with the main difference being the image contrast. The Michelson contrast, defined as $(I_{max}-I_{min})/(I_{max}+I_{min})$, for the line scans were approximately twice larger for WCVD compared to those from the Gaussian 3D. The fast Fourier transform (FFT) and the line scan over the Bragg peaks show that the Bragg peaks are generally more prominent for WCVD as well (see Fig. S10). In addition, the images denoised by Gaussian 3D were generally more blurry than the images from WCVD as shown in the magnified images of Fig. 3.

We additionally evaluate the denoising model performance using the unsupervised mean squared error (uMSE) and unsupervised peak signal-to-noise ratio (uPSNR).[12, 40] Briefly, these metrics assume that experimental images ($y$) are combination of clean signals ($x$) corrupted by independent noise ($w$): $y=x+w$. In such case, the goal of denoising function ($f$) is to remove $w$ by minimizing the mean squared error (MSE) given by $\frac{1}{n}\sum_{i=1}^{n}(x_i - f(y_i))^2$. As the ground truth ($x$) is unknown, we estimate the MSE from the noisy references. For each noisy frame ($y_i$), we form three additional noisy views ($a$, $b$, $c$) of the same scene by using temporally adjacent frames. The key assumption is that these frames have independent, zero-mean noise and the signal is relatively unchanged over the adjacent time frames. Under these assumptions, the mean of $(a - f(y))^2$ contains the denoising error and noise power. This noise power can be obtained from the mean of $\frac{1}{2}(b-c)^2$. In such case, we can obtain the uMSE and uPSNR by

$$\text{uMSE} = \frac{1}{n}\sum_{i=1}^{n}(f(y_i) - a_i)^2 - \frac{1}{2n}\sum_{i=1}^{n}(b_i - c_i)^2 \quad (1)$$

$$\text{uPSNR} = 10\log_{10}\left(\frac{M^2}{\text{uMSE}}\right) \quad (2)$$

where $M$ is the intensity peak of the experimental images. Following the standard practice in image and signal processing, we report uPSNR in dB scale as the power of a signal are generally orders of magnitude larger than the power of the noise.

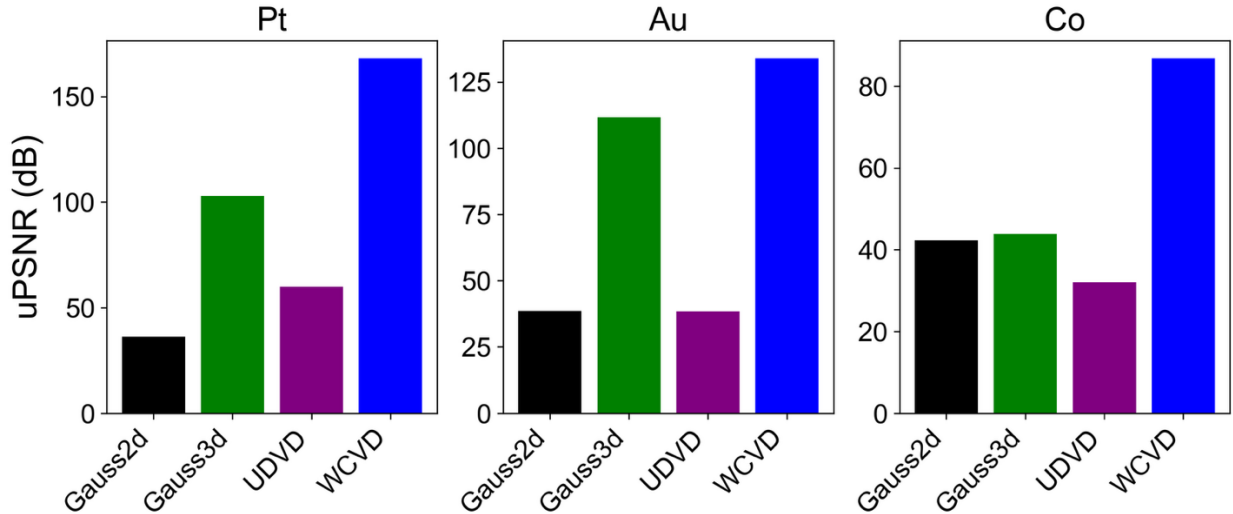

**Figure 4.** Unsupervised peak signal-to-noise ratio (uPSNR) for Pt, Au, and Co datasets for non-ML methods (Gaussian 2D, Gaussian 3D), UDVD, and WCVD.

Figure 4 compares uPSNR for the Pt, Au, Co datasets for the various denoising models utilized in this study. WCVD achieves the highest uPSNR on all datasets, indicating orders-of-magnitude lower residual error after accounting for noise. These results suggest the WCVD model performs particularly well on this metric as it suppresses the noise and enhances image contrast as depicted in Fig. 3. As additional evaluation of the models, we measure the uPSNR for the signal region as depicted in Fig. S11. The results show that UDVD outperforms Gaussian 3D for the Pt dataset and WCVD outperforms all models for all datasets.

## 2.4 Inference speed

The inference speed for the WCVD model was tested for varying number of time frames ($N_t$) and spatial dimensions of the image as $N_x \times N_x$ (Fig. 5). The inference speed was measured on a single NVIDIA A100 GPU by repeating the inference of WCVD model, excluding the pre-processing and post-processing steps, 100 times and getting the average inference speed. The results demonstrate, as expected, that the inference speeds are approximately linearly dependent on the number of pixels of the dataset. Furthermore, the inference speed is in the order of ms per frame, reaching the image speed of conventional GATAN K2 camera. This implies that the model can be applied to in-situ experiments for real time denoising. In comparison, we have compared the inference speed of our model with the UDVD, a ML model for HRTEM images that utilizes

U-Net architecture.[31] The inference speed of UDVD on NVIDIA A100 was 1.78 million pixels per second or 0.59 second per image for 1024×1024 resolution cases. On the other hand, the WCVD inference speed was 454 million pixels per second and 0.00231 second per image for 1024×1024 resolution cases. This comparison shows our light-weight ML architecture employed in this study leads to ~250 times speed up compared to U-Net-based models.

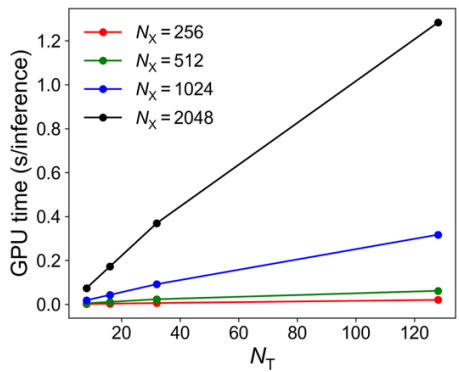

**Figure 5.** Inference speed of WCVD as function of the image size. $N_T$ are the number of snapshots and $N_X$ are the size of the images in one dimension.

## 3. Conclusions

In this study, we have successfully developed and demonstrated a machine learning model based on a self-supervised blind spot-denoising network for HRTEM images. The model has proven to be effectively applicable to diverse experimental conditions, as shown by its performance on out-of-distribution datasets with vastly different pixel intensity distributions, noise levels and resolutions. Our findings indicate that the self-supervised ML model delivers substantial improvements in noise reduction and contrast enhancement of atomic features when compared to traditional non-ML denoising methods as well as U-Net and autoencoder based ML models.

Furthermore, the computational efficiency of the model, with inference speeds in the $O(100)$ frames per second range for typical HRTEM image sizes, confirms its viability for real-time processing during in-situ experiments. This work, therefore, presents a robust and efficient computational tool that can help overcome the inherent noise challenges in low-dose HRTEM, enabling more precise observation and analysis of nanoscale processes at high temporal and spatial resolutions.

## 4. Methods

### 4.1 WCVD model

For the WCVD model, the input data are in-situ HRTEM images with $N_t$ temporal sequences and spatial dimensions of $(N_x, N_y)$ pixels in $x$ and $y$ axes. Such data are organized as tensors with dimensions $(C_{in}, N_t, N_x, N_y)$ where $C_{in}$ is the number of channels representing the pixel intensities. $C_{in}$ is equal to 1 for all HRTEM datasets utilized in this study. As described in section 2.1 and Fig. S1, the denoising ML architecture that consists of a single blind convolutional kernel (BCK) layer followed by multiple layers of $1 \times 1 \times 1$ CNN layers and fast Fourier transform (FFT) operations. The BCK utilizes kernels with temporal and spatial kernel sizes of $(K_t, K_s, K_s)$, with the central kernel weight fixed to 0 to avoid obtaining information about the central pixel. These kernels did not have bias.

Following the BCK, we utilize $N_L$ layers of CNN with kernel size equal to $1 \times 1 \times 1$ that maintain the channel sizes to $C_{out}$. Such architecture is chosen to prevent the pixels from being updated based on the neighboring pixel values. These layers are followed by leaky ReLU activation units with negative slope of 0.1. As the last layer, a $1 \times 1 \times 1$ CNN with $C_{out}$ equal to 1 followed by a ReLU activation is used to output the denoised images.

The BCK converts the input channel size to $C_{out}$. The temporal dimensions utilize valid padding and the output dimensions are reduced by $N_t - K_t + 1$. The spatial dimensions are using periodic padding. The choices of $(K_t, K_s, C_{out}, N_L)$ are hyperparameters used to improve the model performance. We fix $K_t$ to 3, which is the smallest value for a 3D CNN, to maximize the temporal resolution of the model. We have experimented with a range of $N_L$ (1, 2, 4, 8) and $C_{out}$ (1, 2, 3, 4) and chose $C_{out}$ of 3 and $N_L$ of 4 as the hyperparameters for the model based on the denoising performance.

We utilize an ensemble approach where three denoising models are trained simultaneously and their outputs are added with learned weights. The three models were identical in architecture and hyperparameters except for the spatial kernel size ($K_s$), which were chosen to be 3, 5, and 7. This choice was made for two reasons. First, we find the models with larger $K_s$ are more robust to high-frequency noises but less sensitive to the details of the materials, while the smaller $K_s$ leads to contrasting strength and weaknesses. The ensemble model can combine the advantages of both type of kernels. Second, we find $K_s$ dependent artifacts in the FFT at high frequency regime as discussed in SM section 2, which is mitigated by usage of the ensemble model.

After the CNN operations, we apply FFT and mask a circular region at the center. The diameter of circular region is chosen to be $N_s/K_{s,max}$, where $N_s$ is the spatial dimension of the input image and $K_{s,max}$ is the largest spatial kernel size in the ensemble. This choice of mask size is discussed in SM section 2. For the masked FFT results, inverse FFT (IFFT) operation is done to remove the artifacts in the high frequency regime.

For training the model, the loss functions are mean square error (MSE) between the input (with first and last temporal frame discarded to match the size change from valid temporal padding) and the output images. The models were trained 300 epochs with Adam optimizers and learning rate of 0.001. The weight corresponding to central pixel in BCK is initialized to 0. To keep it as zero, we intercept the gradient back propagation process and set the corresponding element in the gradient tensor to 0 before passing the gradient to BCK.

We used the drift correction algorithm by Guizar-Sicarios et al.[41] For low signal-to-noise ratio dataset such as Au, we find that the dataset requires careful denoising before application of the drift correction algorithm. Using the algorithm on the raw dataset led to numerical error. The algorithm also led to noticeable remaining drift when applied to Au dataset after it was denoised by Gaussian 3D filter (see Fig. S12 in the SM). However, WCID and Gaussian with 2D filter led to good drift correction of the Au dataset.

## 4.2 Experimental details for the datasets

To train our model, we used in-situ HRTEM images from 3 sets of experiments (Figure 2, raw). The Pt dataset utilized the published dataset in Figure 1 of Crozier et al.[12, 37] The HRTEM images were observations of Pt nanoparticles on $CeO_2$ at room temperature. The images were taken with Gatan K3 direct electron detector in electron counting mode. The electron dose rate was 2000 $e^-Å^{-2}s^{-1}$ and the frames per second were 75. We used the first 20 frames of the published dataset. More detailed descriptions are given in the main manuscript and supplementary materials of Crozier et al.[12, 37]

The Au and the Co datasets were acquired in the column of $C_s$-Corrected Environmental FEI Titan 80-300 (S)TEM operated at 300kV using Gatan K2-IS direct electron configured in electron counting and linear modes[42], respectively. Images of 5 nm gold nanoparticles (AuNP) functionalized with 3' thiol-functionalized ssDNA following a high ssDNA coverage method were taken with low electron dose at -170°C.[43, 44] Cobalt nanoparticles on $CeO_2$ were taken in 50 mTorr of hydrogen environment at 450°C.[45]

### 4.3 Additional denoising methods

We utilized the Gaussian filtering for non-ML denoising methods. The 3D Gaussian filters were applied using the SciPy package with kernel sizes of (3,3,3). The temporal kernel size of 3 was chosen to match the WCVD model and the spatial kernel size was chosen as larger kernels led to more blurry images. The 2D Gaussian filters were applied using ImageJ.

The ML models utilized Unsupervised Deep Video Denoising Model (UDVD)[12, 31], autoencoder,[38, 46] and Noise2Void.[27] For UDVD, we used the weights publicly available[37] with 'blind-video-net-4' model. Detailed descriptions of the autoencoder models are given in Supplementary Materials section 3. The Noise2Void model utilized the CAREamics package[47] with patch sizes of (64, 64).

# Acknowledgements

This research was supported by the U.S. Department of Energy program "Electron Distillery 2.0: Massive Electron Microscopy Data to Useful Information with AI/ML." This research used Electron Microscopy and Theory and Computation resources of the Center for Functional Nanomaterials (CFN), which is a U.S. Department of Energy Office of Science User Facility, at Brookhaven National Laboratory under Contract No. DE-SC0012704. The Authors would like to thank the authors of Burmistrova et al.[44] and Deng et al.[45] for use of sample images for testing denoising algorithms.

# Author declarations

The authors have no conflicts to disclose.

# Data and code availability

The code and model used in this study are on:

https://github.com/CFN-ETEM/HRTEM_denoising_model

# Supplementary Materials: Machine Learning Pipeline for Denoising Low Signal-To-Noise Ratio and Out-of-Distribution Transmission Electron Microscopy Datasets

Brian Lee[1], Meng Li[1], Judith C. Yang,[1,2,*] Dmitri N. Zakharov,[1,*] Xiaohui Qu[1,*]

[1]Center for Functional Nanomaterials, Brookhaven National Laboratory, Upton, NY, United States
[2]Department of Chemical and Petroleum Engineering, University of Pittsburgh, Pittsburgh, PA, United States
*Corresponding author email: jyang1@bnl.gov, dzakharov@bnl.gov, xiaqu@bnl.gov

## 1. Weight-centric video denoising model architecture

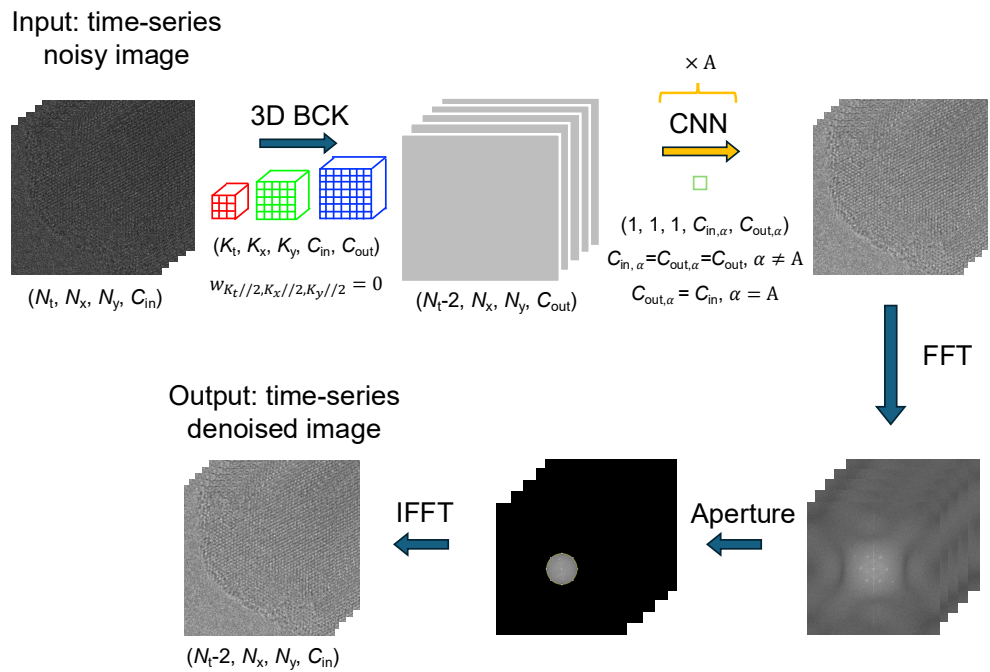

**Figure S1. Model architecture of weight-centric video denoising model.** Time-series TEM images of ($N_t$, $N_x$, $N_y$, $C_{in}$) dimensions are used as input. An ensemble of blind convolution kernels (BCK) followed by 1×1×1 convolution neural network (CNN) kernels are used to output denoised image. Finally, FFT operation, masking, and IFFT operation are performed

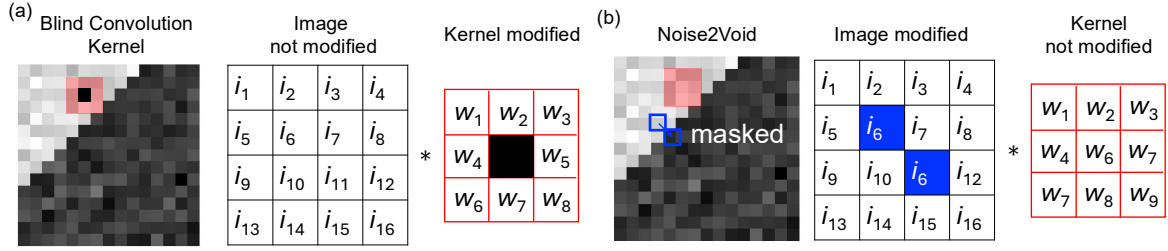

**Figure S2. Illustration of BCK and Noise2Void.** (a) The image inputs are not modified for BCK while the central pixel of the kernel weights is masked. (b) For Noise2Void, the kernel weights of CNN are not modified. Instead, a subset of input pixels is chosen to be replaced by neighboring pixels and the model is trained to output the original input pixels.

## 2. Fast Fourier Transform Analysis

The fast Fourier transform (FFT) analyses of Au dataset images are depicted in Fig. S3. In contrast to the FFT of the raw images (Fig. S3c), the Fourier spectra obtained after Gaussian filtered (Fig. S3a) and WCVD denoising (Fig. S3b) display pronounced, alternating bands of high and low power in the high-frequency domain. For the Gaussian filtered images, we observe a square, bright region in the low-frequency regime surrounded by dark regions as well as horizontal and vertical streaks. This is because Gaussian filter is a low-pass filter that smooths images by reducing the sharp details in high-frequency region. Mathematically, the FFT of Gaussian filtered images can be defined as,

$$F\{\hat{I}\}(u,v) = F\{I\}(u,v)F\{G\}(u,v). \tag{1}$$

where $F, G$ denote the FFT and Gaussian filtering. $\hat{I}$ and $I$ represent the Gaussian filtered images and raw images. The Gaussian kernel is defined as $G(x, y, \sigma) = \frac{1}{2\pi\sigma^2} e^{-\frac{x^2+y^2}{2\sigma^2}}$ and its Fourier transform is defined as $\hat{G}(x, y, \sigma) = e^{-2\pi^2\sigma^2(u^2+v^2)}$. This signifies that the width of bright region for Gaussian filtered images is proportional to $\sim 1/\sigma$ with respect to the radius $r = \sqrt{u^2 + v^2}$ in frequency space.

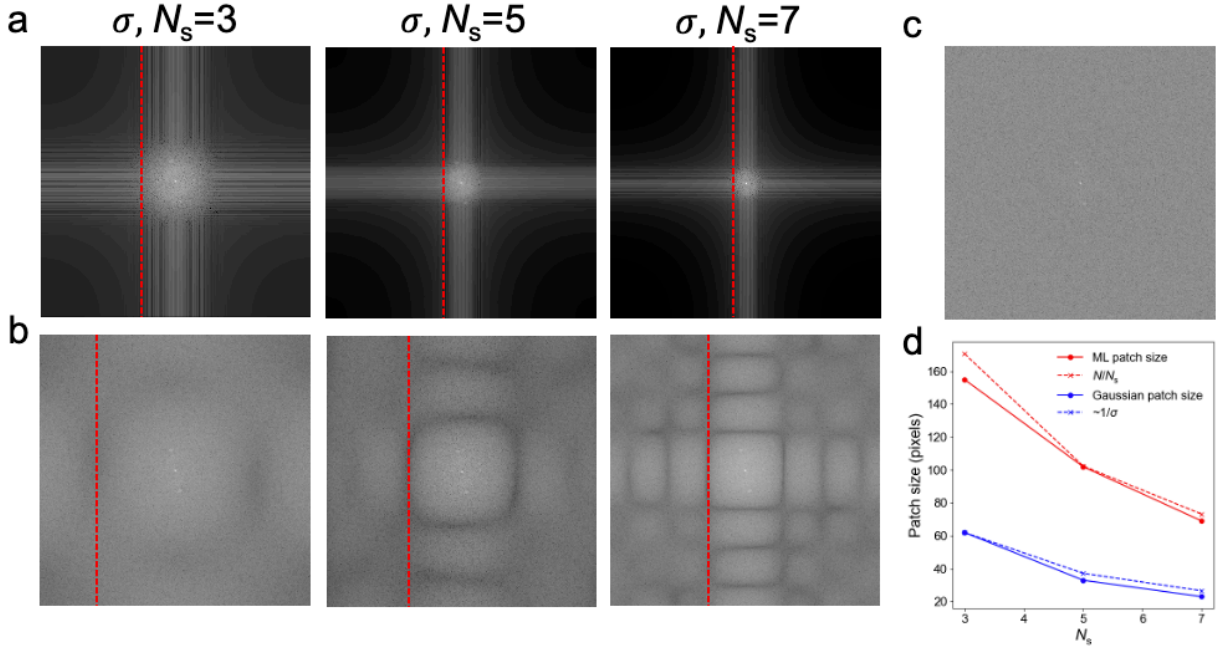

**Figure S3.** FFT of Au sample from (a) Gaussian filtered, (b) Blind convolution kernel denoised, and (c) raw images. The Gaussian filter size ($\sigma$) and BCK spatial kernel sizes ($N_s$) are noted on the images. Results from varying spatial kernel sizes ($\sigma$, $N_s$) of 3, 5, 7 are shown for Gaussian-filtered and ML denoised images. (d) The pixel positions of the dark regions of the FFT magnitudes as depicted by red dashed lines on (a) and (b).

This is demonstrated in the size of the bright central regions as functions of spatial kernel size ($\sigma$) depicted in Fig. S3d. Here, we are plotting the radius in frequency space shown by red dashed lines on Fig. S3a. We find that the values decay proportional to the $\sim 1/\sigma$.

For the ML, we observe again bright low-frequency region surrounded by nearly square, dark regions. For the spatial kernel sizes ($N_s$) of 5 and 7, we see repetition of these dark regions in checkerboard patterns. This is because the convolution neural network (CNN) is using information within square kernels and tiling these kernels across the image. This is analogous to convolution of box functions, whose Fourier transform leads to sinc functions. This results in the magnitude of FFT having periodically reduced values at intervals of $1/N_s$. As depicted in Fig. S3d, the positions of the first dark region occur at $N/N_s$ distance from the origin in the frequency space where $N$ is the size of the images. Consequently, spatial frequencies corresponding to real-space periods finer than $d_{min} \approx N_s$ pixels are strongly attenuated, setting the upper bound of Nyquist resolution achievable with the current CNN architecture.

# 3. Autoencoder for denoising

We trained a convolutional autoencoder to denoise the HRTEM images. The inputs to the model were 2D arrays of HRTEM images. The model followed an encoder-decoder structure with symmetric convolutional blocks. The encoder consisted of sequential convolutional blocks, each comprising two 3 × 3 convolution with batch normalization and ReLU activation, followed by 2 × 2 max pooling layers. The number of channels was doubled after each downsampling step, beginning with 32 filters in the first block. The bottleneck layer was applied after the final downsampling step as a convolutional layer with same number of channels as the final downsampling step. The decoder involved layers of upsampling blocks consisting of a deconvolutional network[1] and a convolutional network both followed by batch normalization and ReLU activation. The target output of the model was the input and MSE loss was utilized. Similar to the other ML models, the training data was Pt and the model was tested on Pt, Au, and Co datasets.

We trained models with varying number of downsampling and upsampling blocks ($N_{block}$) and the results are depicted in Fig. S4. For Pt dataset, the autoencoder models maintains clarity of atomic columns but has higher variance in the gas area for $N_{block}=2$ and checkerboard type of artifacts for $N_{block}=4$. For the Au datasets, the $N_{block}$ of 2 and 3 lead to barely visible atomic features. $N_{block}$ of 4 leads to images that resemble a nanoparticle but do not show the clear lattice structures observed in WCVD results. For Co datasets, $N_{block}$ of 2 leads to reasonable images with clear atomic features while the features of images from models with higher $N_{block}$ are not as distinct or contains significant checkerboard artifact. Overall, the autoencoders display some denoising ability, but require varying architectures ($N_{block}$) for different datasets and the denoising performance is inferior to WCVD model or 3D Gaussian filtering.

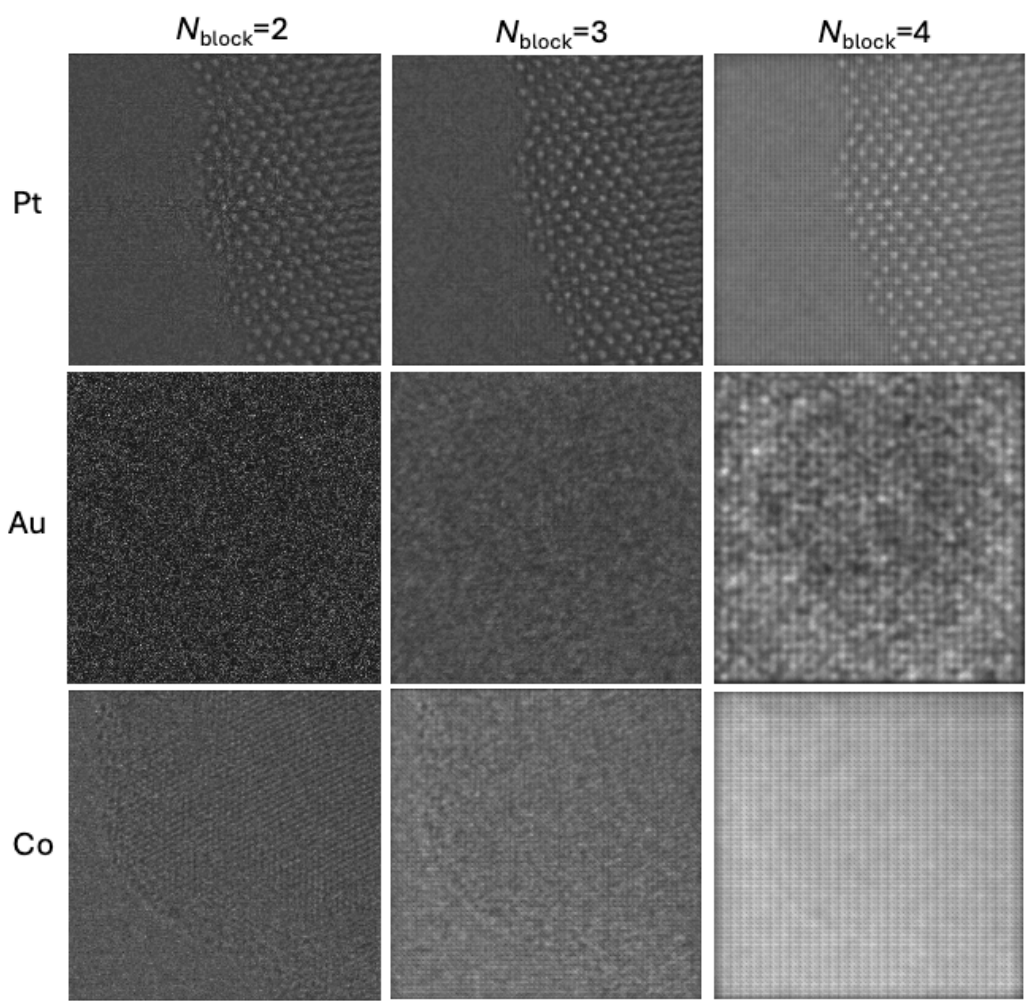

**Figure S4.** Images denoised from autoencoder with varying downsampling and upsampling blocks.

# 4. Histogram of pixel intensities

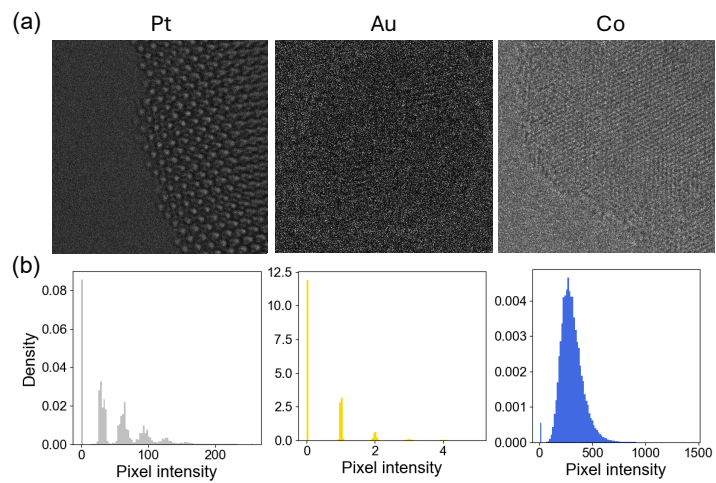

**Figure S5.** (a) Raw images and (b) pixel intensity histograms of Pt, Au, and Co datasets. Images are the first slices of each datasets while the histograms are for the entire datasets.

# 5. Denoised images of Pt, Au, and Co

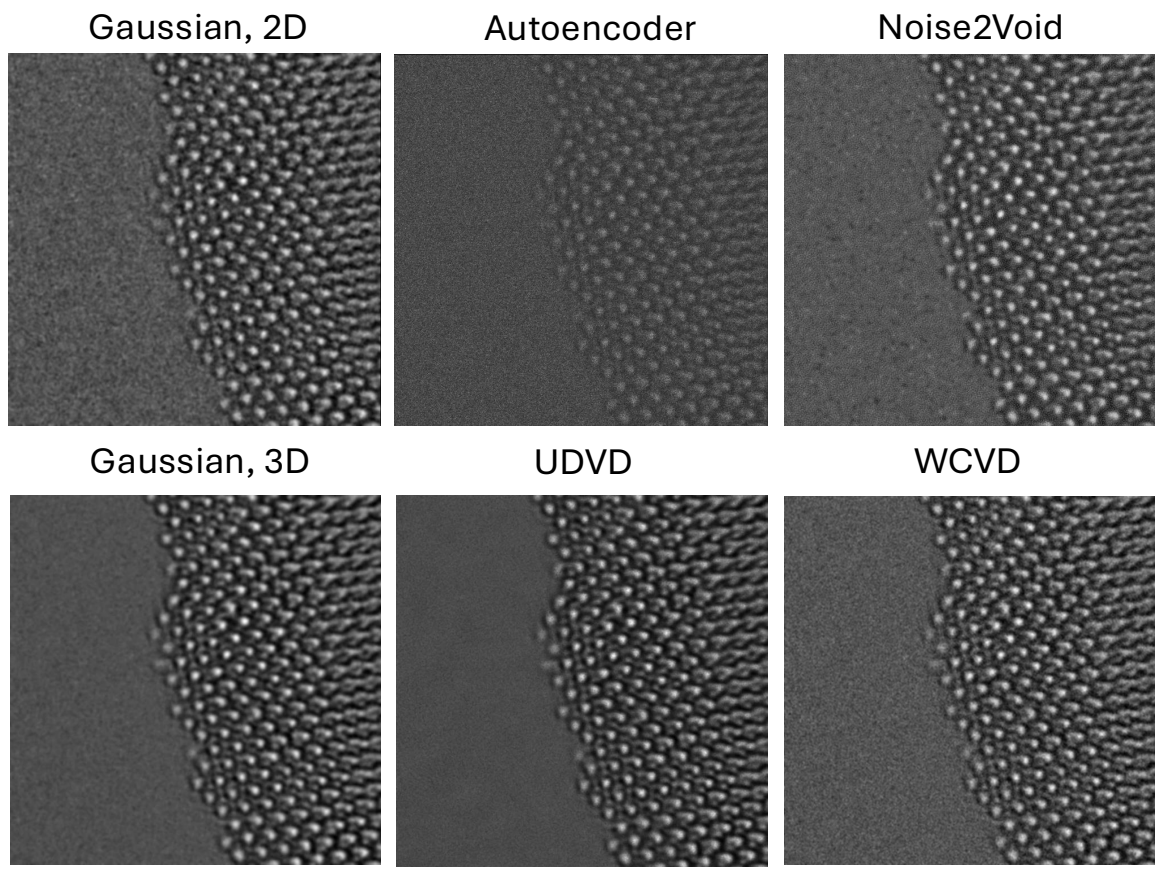

**Figure S6.** Denoised images of Pt depicted in Fig. 2 of the main manuscript.

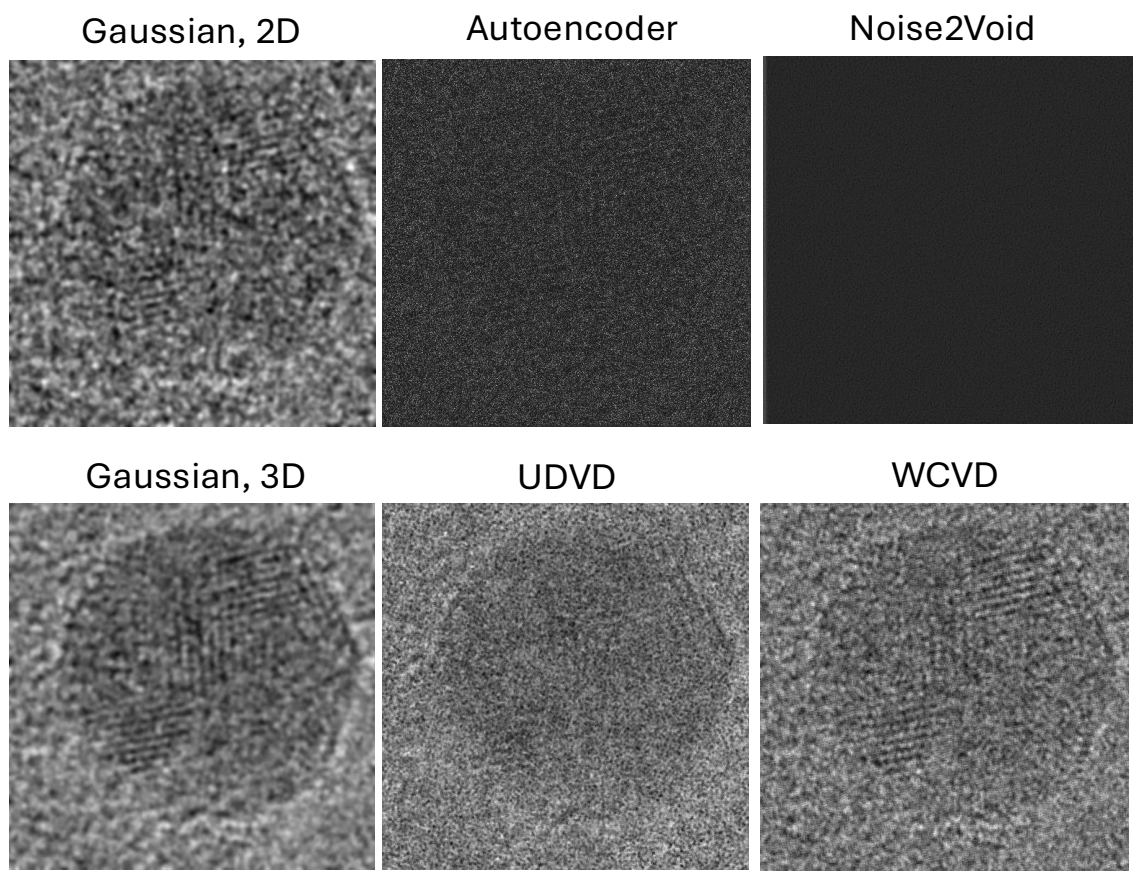

**Figure S7.** Denoised images of Au depicted in Fig. 2 of the main manuscript.

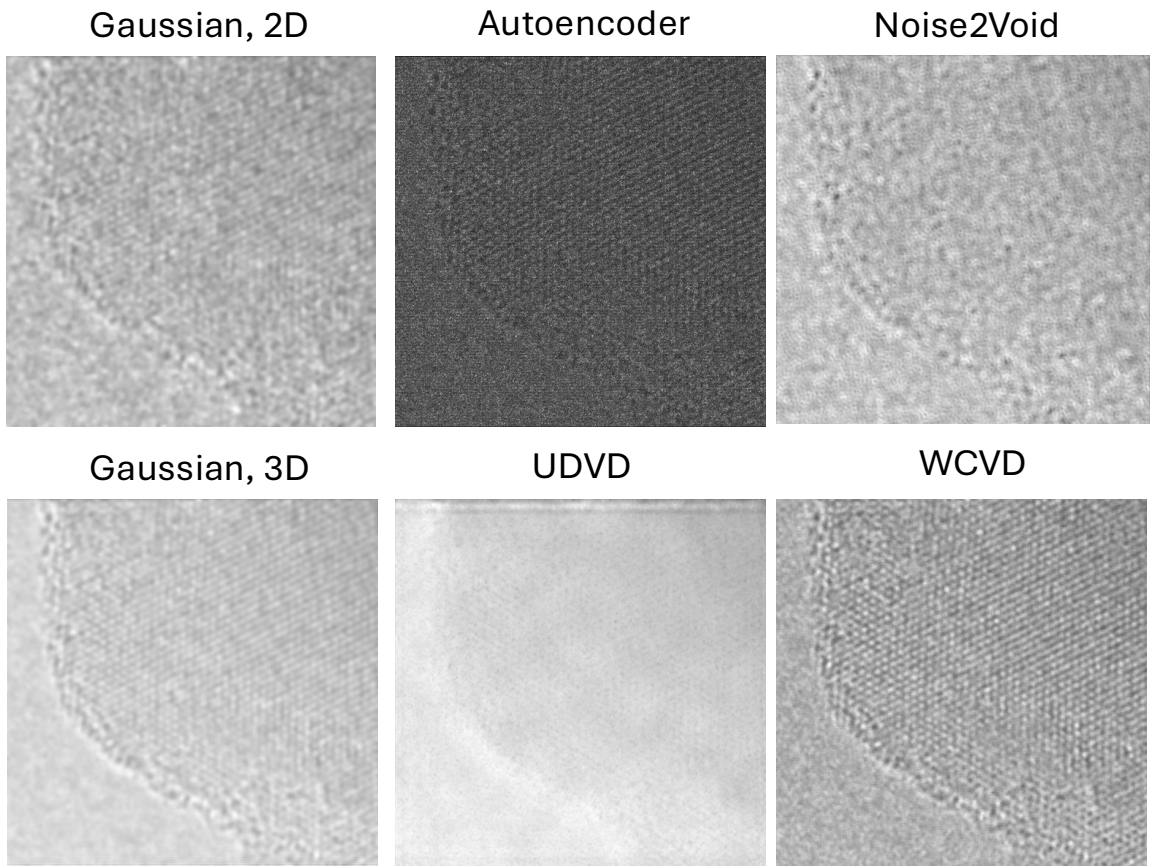

**Figure S8.** Denoised images of Co depicted in Fig. 2 of the main manuscript.

# 6. Additional results on magnified images

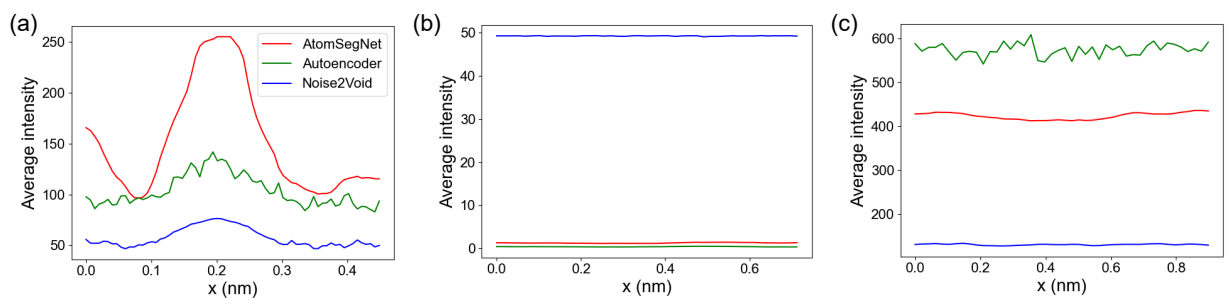

**Figure S9.** Line scan over atomic columns of (a) Pt, (b) Au, and (c) Co samples for AtomSegNet, Autoencoder, and Noise2Void.

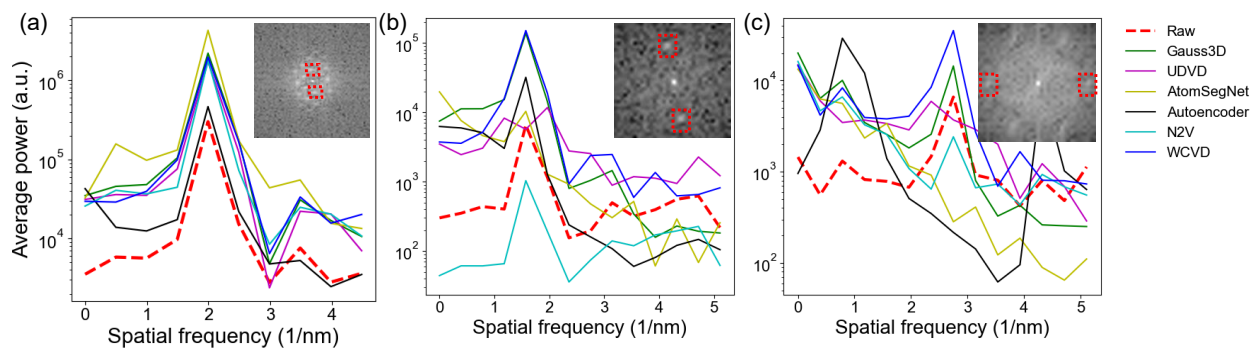

**Figure S10.** Line scan over Bragg peaks from FFT of (a) Pt, (b) Au, and (c) Co samples for raw, Gaussian3D, UDVD, AtomSegNet, Autoencoder, Noise2Void, and WCVD.

# 7. Unsupervised Peak Signal-to-Noise Ratio

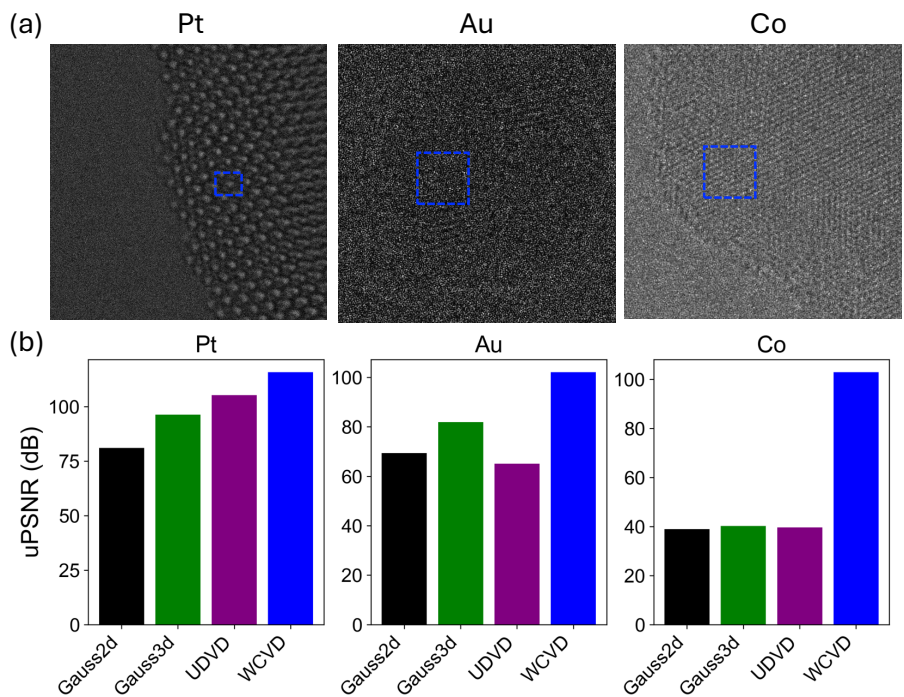

**Figure S11.** (a) Raw data of Pt, Au, and Co. Blue rectangles represent the chosen signal region. **(b)** uPSNR in the signal region.

# 8. Drift correction

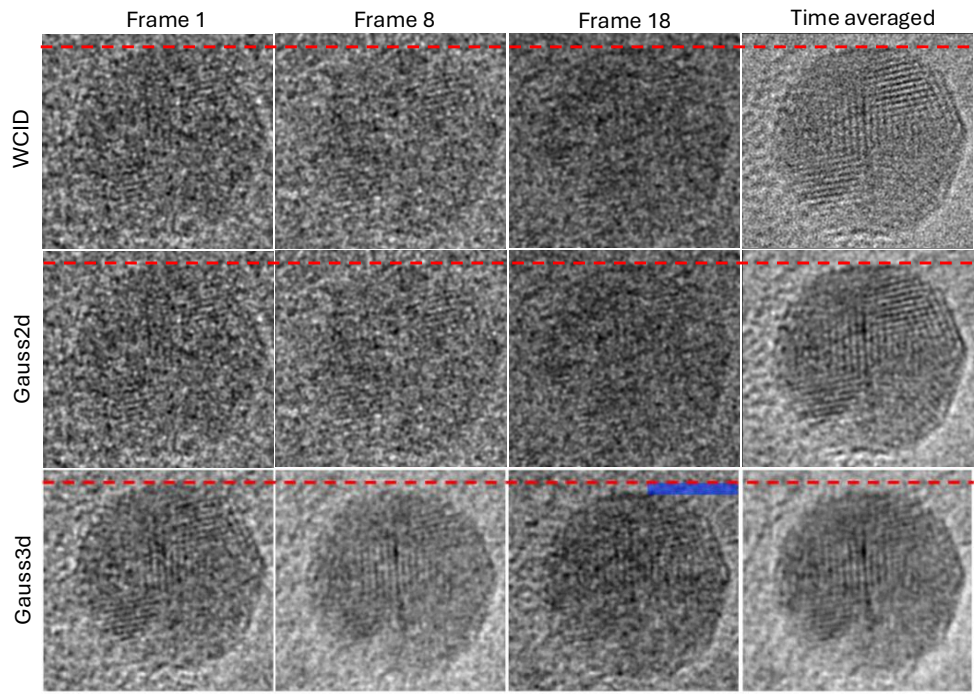

**Figure S12.** Drift corrected images of the Au dataset after WCID, Gaussian 2D, and Gaussian 3D filters were applied to the raw Au dataset.

# Reference

(1) Zeiler, M. D.; Krishnan, D.; Taylor, G. W.; Fergus, R. Deconvolutional networks. In *2010 IEEE Computer Society Conference on Computer Vision and Pattern Recognition*, 13-18 June 2010, 2010; pp 2528-2535. DOI: 10.1109/CVPR.2010.5539957.